\documentclass[preprint,12pt]{elsarticle}

\usepackage{graphics}
\usepackage{epsfig}

\usepackage{amsmath}
\usepackage{amssymb}

\usepackage[dvips]{color}
\usepackage{colordvi}
\usepackage{bm}
\usepackage{graphics}
\def\beq{\begin{equation}}
\def\be{\begin{eqnarray}}
\def\eeq{\end{equation}}
\def\ee{\end{eqnarray}}

\def\gsim{\buildrel > \over {_{\sim}}}

\journal{Physics Letters B}

\begin{document}

\begin{frontmatter}



\title{Nuclear effects in neutral current quasi-elastic neutrino interactions}


\author[label1,label2]{Omar Benhar}
\author[label2]{Giovanni Veneziano}

\address[label1]{INFN, Sezione di Roma, I-00185 Roma, Italy}
\address[label2]{Dipartimento di Fisica, ``Sapienza'' Universit\`a di Roma, 
I-00185 Roma, Italy}

\begin{abstract}
The interpretation of the charged (CCQE) and neutral (NCE) current quasi elastic events collected by 
the MiniBooNE collaboration involves a number of unresolved issues. While it has been 
suggested that the data can be explained in terms of an effective nucleon axial mass, $M_A$,
the results of our theoretical calculations suggest that the CCQE and NCE samples cannot be 
described by the same value of $M_A$. We argue that the disagreement between theory and data 
may arise from the uncertainties associated with the flux average procedure. We also analyze the 
role of the strange quark in NCE interactions and find that, due to a cancellation between 
proton and neutron contributions, it turns out to be negligible.
\end{abstract}


\end{frontmatter}



\section{Introduction}
\label{intro}

The MiniBooNE collaboration has recently collected an extensive data set of quasielastic
neutrino nucleus scattering events, in both the charged-current (CCQE) \cite{BooNECCQE} and
neutral current (NCE) \cite{BooNENCE} channels, using a Carbon target.

In the CCQE channel, quasielastic neutrino-nucleon interactions are described in terms of the vector form
factors  $F_1^{p,n}(Q^2)$ and $F_2^{p,n}(Q^2)$ ($Q^2= -q^2$, $q$ being the four-momentum
transfer, while the superscripts $p$ and $n$ correspond to proton and neutron, respectively), that have been 
precisely measured in electron-proton and electron-deuteron scattering
experiments \cite{VFF},
and the axial form factor $F_A(Q^2)$ \cite{bernard,bodek,nomad}. In addition, NCE interactions 
are also affected by the form factors $F_1^s$, $F_2^s$ and $F_A^s$,  arising from strange quark
contributions \cite{HAPPEX,Horowitz,Alberico,Liu}. The results of recent experiments \cite{HAPPEX}
indicate that  $F_1^s$, $F_2^s$ are vanishing, whereas the axial form factors  $F_A$ and $F_A^s$ are
assumed to be of dipole form, and their $Q^2$-dependence is parametrized in terms of the axial mass $M_A$.

The measured cross sections turn out to be consistently larger than the predictions of Monte Carlo simulations
carried out using the relativistic Fermi gas (RFG) model of the nucleus and the value of the axial mass resulting
from the world average of the deuterium data, $M_A = 1.03$ GeV \cite{nomad}.
In order to bring the predictions of the RFG model into agreement with the data, the authors of 
Refs.\cite{BooNECCQE,BooNENCE} use a significantly larger value of the axial mass, $M_A \gsim 1.35$ GeV, and 
introduce the additional parameter $\kappa$, meant to improve the treatment of Pauli blocking. 
The K2K collaboration also reported a large value of the axial mass, $M_A \sim 1.2$ GeV, resulting form the analysis of its sample of CCQE  
events \cite{K2K}. Moreover, the best fit to the neutral current data is obtained using a non vanishing strange quark contribution $\Delta s$, 
determining the value of $F_A^s$ at $Q^2=0$ \cite{BooNENCE}. 

It has been suggested that the large value of $M_A$ may be regarded as an {\em effective} axial mass, modified by 
nuclear effects not taken into account in the RFG model \cite{BooNECCQE}. However, the results obtained  
using more advanced models appear to rule out this explanation. In fact, numerical calculations carried out 
using realistic nuclear spectral functions, extensively employed in the analysis of electron-nucleus scattering data 
\cite{RMP}, indicate that reproducing  the CCQE measured cross sections requires an even larger value of 
$M_A$ \cite{axmass,newpar}. 

The purpose of this work is the extension of the spectral function approach of Refs.\cite{axmass,newpar} to the description of 
NCE interactions and the quantitative analysis of the $M_A$ and $\Delta s$ dependence of the resulting cross sections.
The main elements of our approach are outlined in Section \ref{formalism}, while the numerical results are discussed in Section \ref{res}.
Finally, in Section \ref{summary} we summarize our findings and state the conclusions.

\section{Formalism}
\label{formalism}

We consider the neutral current process
\beq\nu_\mu + ^{12}\mkern -5mu C \rightarrow \nu_\mu + X \ ,
\eeq
in which a neutrino carrying initial four-momentum $k = (E_\nu,\bf{k})$ scatters off a Carbon target to a state of four-momentum $k' = (E'_\nu,\bf{k'})$, the target final state being undetected. In the impulse approximation (IA) scheme \cite{PRD}, stating that when the magnitude of the momentum transfer 
$|\bf{q}|$ is large enough 
(i) the target nucleus is seen by the probe as a collection of individual nucleons and (ii) in the final state the knocked out nucleon and the recoiling nucleus evolve independently of one another, the differential cross section can be written in the form
\beq
d\sigma_{IA} = \int d^3p\;dE\; P(\mathbf{p},E)\; d\sigma_{elem} \ ,
\eeq
where $d\sigma_{elem}$ is the neutrino-nucleon cross section and $P(\bf{p},E)$ is the spectral function of the target nucleus, 
yielding the probability distribution of finding a nucleon of momentum $\bf{p}$ and removal energy $E$ in the nuclear target.

\subsection{Neutrino nucleon cross section}The NCE neutrino nucleon cross section in the center of mass frame reads
\beq
\frac{d\sigma_{elem}}{d\Omega} = \frac{|\bar {\mathcal M}|^2}{64 \pi^2 (E_\nu + E_p)^2} \left( \frac{E'}{E_\nu} \right ) ,
\eeq
where $E_p$ is the nucleon energy and $|\bar {\cal M} |$ is Feynman's invariant amplitude, averaged over the spins of the initial state
 particles and summed over the spins of the particles in the  final state. 

Feynman's amplitude can be written as
\beq
{\mathcal{M}}= \frac{i}{2\sqrt2} G_F \:\underset{leptonic\;current}{\underbrace{ \bar\nu(k^{'})\gamma_\mu(1- \gamma_5)\nu(k)}}\;\;\underset{hadronic\;current }{\underbrace{<N(p^{'})|J_Z^\mu|N(p)>}},\eeq
where $\nu(k)$ and $\bar \nu(k')$ are the Dirac spinors associated with the initial and final state neutrino, respectively, the kets $| N(p)>$ and $| N(p')>$ represent the initial and final nucleon state, and $J_Z$ is the hadronic neutral current.
While the leptonic current has a simple V-A structure, completely determined by the leptons kinematics, the hadronic current 
is more complex, on account of the strong interactions occurring between the nucleon constituents.

The hadronic neutral weak current can be written in the general form
\beq \begin{split}
&<N(p^{'})|  J_Z^\mu|N(p)> = \\ & <N'|\left [\gamma^\mu F_1^z(Q^2)+ \frac{i\sigma^{\mu\nu}q_\nu}{2 M}F_2^z(Q^2)+ \gamma^\mu \gamma^5 F_A^z(Q^2)\right ]|N> ,
\end{split}
\eeq
where $F_1^z(Q^2)$, $F_2^z(Q^2)$ and $F_A^z(Q^2)$ are the Dirac, Pauli and axial form factors for neutral current interactions, respectively,  
taking into account the strange quark content of the nucleon.

The form of the neutral weak current  
\beq
J_Z = J_3 - 2sin^2\theta_W J_{em} \ ,
\eeq
where  $J_3$, $J_{em}$ and $\theta_W$ are the third component of the isospin current, the electromagnetic current and Weinberg's angle, respectively, 
suggests the following parametrization of the form factors
\beq
\begin{split}
F_1^{z,p}(Q^2)& =  \frac{1}{2}\left [ \bar F_1(Q^2) - F_1^s(Q^2) \right ] - 2 sin^2 \theta_W F_1^{p}(Q^2), \\
F_1^{z,n}(Q^2)& =  \frac{1}{2}\left [ - \bar F_1(Q^2) -F_1^s(Q^2) \right ] - 2 sin^2 \theta_W F_1^{n}(Q^2) , \\
F_2^{z,p}(Q^2)& = \frac{1}{2} \left [ \bar F_2(Q^2) -F_2^s(Q^2) \right ]- 2 sin^2 \theta_W F_2^{p}(Q^2)  , \\
F_2^{z,n}(Q^2)& = \frac{1}{2} \left [ - \bar F_2(Q^2) -F_2^s(Q^2) \right ] - 2 sin^2 \theta_W F_2^{n}(Q^2) , \\
F_A^{z,p}(Q^2)& = \frac{1}{2}F_A(Q^2) -\frac{1}{2}F_A^s(Q^2)  , \\
F_A^{z,n}(Q^2)& = -\frac{1}{2}F_A(Q^2) -\frac{1}{2}F_A^s(Q^2)  ,
\end{split}
\eeq
where
\beq
\bar F_i(Q^2) = F_i^{p}(Q^2) - F_i^{n}(Q^2) \;\:\:\:\:\: \:\;\;\;\:\;\; i = 1,2 \ ,
\eeq
and $s$ indicates the strange quark contribution. As stated above, $F_1^s$ and $F_2^s$ are vanishing, while $F_A^s$ is assumed to 
have a dipole $Q^2$ dependence
\beq
F_A^s(Q^2) = \frac{\Delta s}{\left ( 1 + \frac{Q^2}{M_A^2} \right )^2} \; ,
\eeq
$\Delta s$ being the strange quark contribution to the nucleon spin at $Q^2 = 0$.

Following Ref. \cite{Horowitz}, we parametrize Feynman's amplitude $ {\cal M}$ in terms of six contributions according to
\beq
|\bar{{\mathcal M}}|^2 = 4 G_F^2 (V_{11}+V_{12}+V_{22}+A+V_{A1}+V_{A2}) \; , 
\eeq
with
\beq
\begin{split}
V_{11} & =  4 (F_1^z)^2 \left [ p \cdot kk' \cdot p' + p' \cdot kk'\cdot p - M^2 k\cdot k' \right]  ,\\
V_{12} & =  - 4 F_1^z F_2^z \; k \cdot k' (p' - p) \cdot (k - k')  ,\\
V_{22} & =  \frac{2(F_2^z)^2}{M^2} k \cdot k' \left [ p \cdot kp'\cdot k + p \cdot k' p'\cdot k' +  M^2 k\cdot k'   \right ]     ,\\
A  &=  4 (G_A)^2 \left [ p \cdot k k' \cdot p' + p' \cdot k k'\cdot p + M^2 k\cdot k'       \right ]        ,\\
V_{A1} &=   8 G_A F_1^z \left [ p \cdot k p' \cdot k' - k \cdot p' p \cdot k'  \right ]   ,\\
V_{A2} &=     4 G_A F_2^z \; k \cdot k'(k+k') \cdot (p+p') ,
\end{split}
\eeq
where $M$ is the nucleon mass and $G_A = - F_A^z (g_A - \Delta s)/(g_A + \Delta s)$, with $g_A = F_A(Q^2 = 0)$.

\subsection{Target spectral function}

Accurate {\em ab initio} calculations of the spectral function $P( \bf{p},E)$, based  on realistic nuclear hamiltonians, can only be carried out 
for the lightest nuclei ($A \le 4$) \cite{B63,C21,A395,C41,86,PRC47} and in the limit of uniform nuclear matter ($A \rightarrow \infty$) \cite{BFF,Ramos}. 
In the case of medium-heavy nuclei, the calculation of $P( \bf{p},E)$ involves severe difficulties,  and one has to resort to some 
simplifying assumptions.

Within the RFG model \cite{Smith,Moniz} the nucleus is described as a degenerate gas of non-interacting nucleons. 
According to this picture the spectral function takes the simple form
\beq \label{p_rfgm}
P_{RFGM}(\mathbf{p},E) = \left ( \frac{6 \pi^2 A}{p_F^3} \right ) \Theta(p_F - \mathbf{p}) \; \delta(E_{\mathbf{p}} - E_B + E) \; ,
\eeq
where $E_{\mathbf{p}} = \sqrt{M^2 + |{\bf p}|^2}$ is the energy of a free nucleon carrying momentum ${\bf p}$. The Fermi momentum $p_F$ and the average binding energy $E_B$ are the model parameters,  to be adjusted to reproduce the experimental data. 

The spectral function in Eq. \ref{p_rfgm} is non vanishing only at  $|\mathbf{p}| < p_F$. However,  electron-nucleus scattering 
experiments have provided unambiguous evidence of strong nucleon-nucleon correlations, that give rise to virtual scattering processes leading to the excitation of nucleons to states of large momentum and removal energy \cite{RMP}. Hence, the quantitative analysis of neutrino-nucleus interactions 
requires a more realistic spectral function, taking into account correlation effects. 

In our work we have used the Carbon spectral function of Ref.\cite{PkE}, obtained within the Local Density Approximation (LDA) combining the information extracted from measurements of the coincidence $(e,e^\prime p)$ cross section with theoretical calculations of the 
spectral function of uniform nuclear matter at different densities.

The resulting $P(\mathbf{p},E)$ consists of two contributions \cite{PkE}
\beq
P_{LDA}(\mathbf{p},E) = P_{MF}(\mathbf{p},E) + P_{corr}(\mathbf{p},E) \; ,
\eeq
arising from the nuclear mean field and from nucleon-nucleon correlation.

The mean field spectral function reads
\beq
P_{MF}(\mathbf{p},E) = \sum_{n} Z_n |\phi_n(\mathbf{p})|^2 F_n(E - E_n) \; ,
\eeq
In the above equation, $\phi_n(\mathbf{p})$ is the squared momentum-space wave function of the $n$-th shell model state, whose width is described by the 
Lorentzian $ F_n(E - E_n)$, $Z_n$ is the corresponding spectroscopic factor and the sum extends to all states belonging to the Fermi sea.
In the absence of correlation $Z_n \rightarrow 1$ and $ F_n(E - E_n) \rightarrow \delta(E - E_n)$.

The correlation contribution to the LDA spectral function can be written in the form
\beq
P_{corr}(\mathbf{p},E)=\int d^3 r \;\rho_A(\mathbf{r})\;P^{NM}_{corr}(\mathbf{p},E;\rho=\rho_A(\mathbf{r}))\; ,
\eeq
where $\rho_A(\mathbf{r})$ is the nuclear density profile and $P^{NM}_{corr}(\mathbf{p},E;\rho)$ is the correlation part of
the nuclear matter spectral function at density $\rho$, whose calculation is described in Ref. \cite{PkE}. 
Correlation effects turn out to be sizable, leading $\sim$  20\% of the strength to the region of large momentum ($|{\mathbf p}| > p_F$) and large energy \cite{PkE}.


In the IA scheme statistical correlations leading to the suppression of the phase-space available to the final state nucleon, generally referred to as Pauli Blocking (PB), 
are not taken into account. In order to introduce their effect in our calculations, we have modified the spectral function according to \cite{PRD}
\beq
P(\mathbf{p},E)\; \Rightarrow \; P(\mathbf{p},E) \; \Theta (| \mathbf{p}+\mathbf{q}|- \bar{p}_F) \; ,
\eeq
where ${\mathbf q}$ is the momentum transfer and $\bar{p}_{F}$ is the average Fermi momentum of the nucleus,  defined as
\beq 
\label{aaaa1}
\bar{p}_F = \int d^3 r \rho_A(\mathbf{r})p_F(\mathbf{r}) \; , 
\eeq
with
\beq
p_F(\mathbf{r}) = \left( \frac{3\pi^2\rho_A(\mathbf{r})}{2} \right)^{1/3}\;.
\label{aaaa2}
\eeq
For a Carbon target, Eqs. (\ref{aaaa1}) and  (\ref{aaaa2}) lead to $\bar{p}_F = 225$ MeV.
The inclusion of PB, while leaving unaffected the cross sections at large $Q^2$, leads to an appreciable quenching in the region 
of low $Q^2$ \cite{PRD}.

\section{Results}
\label{res}

\begin{figure}
\includegraphics[scale=.4]{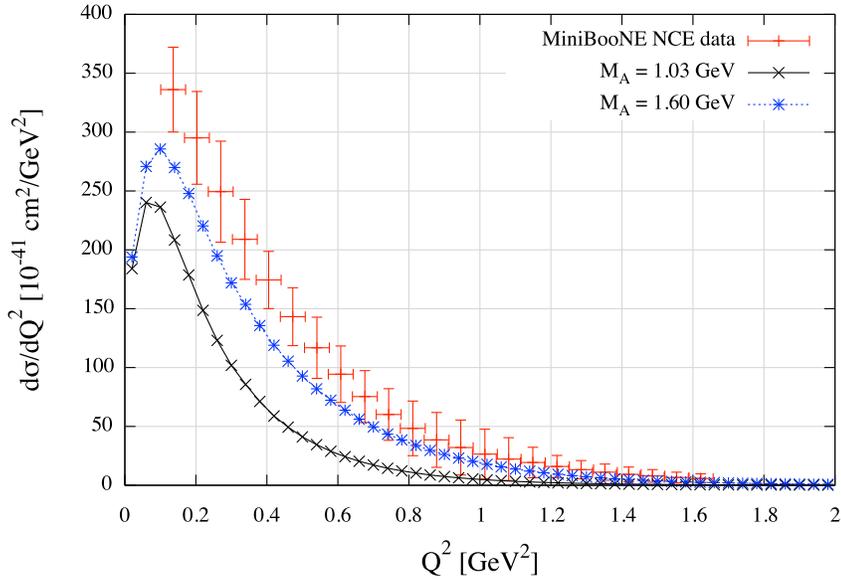}
\caption{NCE flux averaged $Q^2$-distribution for different values of the axial mass. The data points are taken from Ref. \cite{BooNENCE}}
\label{fig1}
\end{figure}

We have computed the $Q^2$-distribution, averaged over the MiniBooNE flux, using the Carbon spectral function of Ref. \cite{PkE}.
Figure \ref{fig1} shows the results corresponding to different values of the axial mass and $\Delta s = 0$, compared
to the experimental data of Ref. \cite{BooNENCE}.
It clearly appears that the value of the axial mass yielding a good fit of the MiniBooNE CCQE distribution,  $M_A = 1.6$ GeV  \cite{newpar},  does not reproduce 
NCE data.

\begin{figure}
\includegraphics[scale=.4]{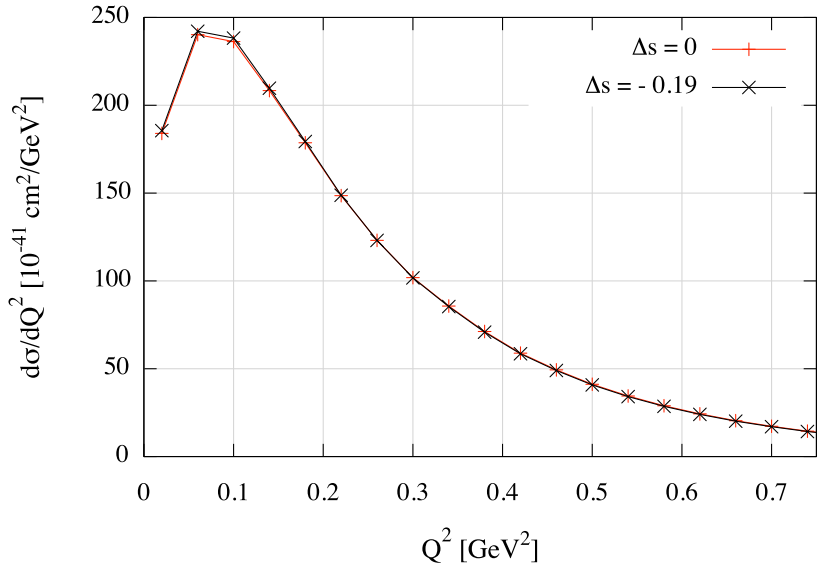}
\caption{NCE flux averaged $Q^2$-distribution for different values of $\Delta s$.}
\label{fig2}
\end{figure}

To illustrate the dependence of our results on $\Delta s$, in Fig. \ref{fig2} we show the flux averaged $Q^2$-distribution for neutrinos interacting with a Carbon 
target, obtained using $M_A=1.03$ GeV and setting $\Delta s = 0$ and $\Delta s = -0.19$, the latter being the lowest value that can be found in the literature 
\cite{Alberico,Liu,BNL}. 
  
It is apparent that the distribution is nearly independent of $\Delta s$. As a consequence, the results displayed in Fig. \ref{fig1} have been 
obtained neglecting the strange quark contribution to the axial form factor. 

\begin{figure}
\includegraphics[scale=.4]{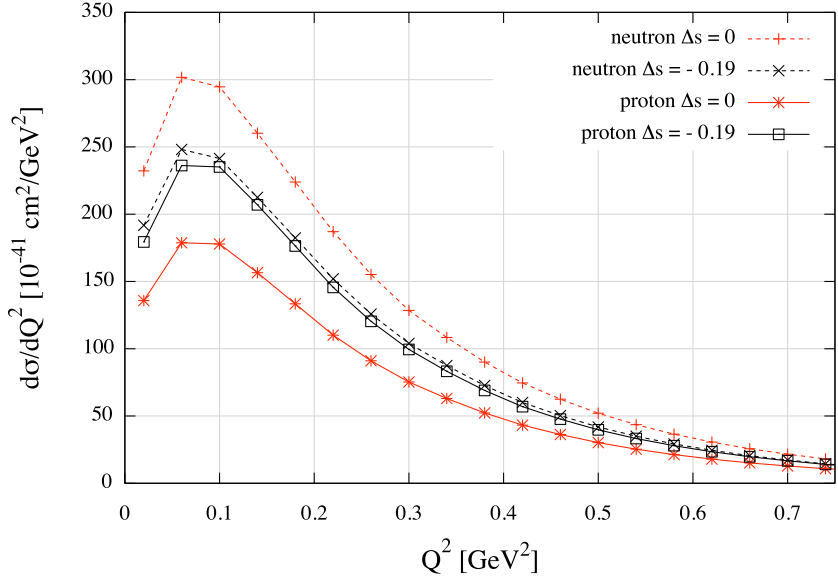}
\caption{NCE proton and neutron contributions to the Carbon $Q^2$-distribution.}
\label{fig3}
\end{figure}

In order to analyze the difference between the proton and neutron contributions,  in Fig. \ref{fig3} we show 
the same distributions as in Fig. \ref{fig2}, calculated for a $A=12$ target consisting of protons or neutrons only.
It turns out that the strange quark produces a suppression of the neutron contributions and an enhancement of the proton 
contribution of about the same size. As a result, the two effects largely cancel each other in the Carbon $Q^2$-distribution, as seen in Fig. \ref{fig2}.

The role of strange quarks had been previously discussed in Refs. \cite{Horowitz,Garvey}, whose authors proposed to determine $\Delta s$ 
from the ratio
\beq
\left( \frac{d \sigma}{dQ^2} \right)_{neutron}^{NCE} / \left( \frac{d \sigma}{dQ^2} \right)_{proton}^{NCE} \ ,
\eeq
that does not suffer from the uncertainties associated with the incoming neutrino flux and is very sensitive to variations of $\Delta s$.

  
\section{Conclusions}
\label{summary}


Our work indicates that the theoretical analysis of the MiniBooNE NCE data sample involves the same 
difficulties already emerged in the studies of CCQE  interactions \cite{newpar}.

The results discussed in Section \ref{res}, showing that  it is impossible to describe both the CCQE and NCE data sets using the same 
value of the axial mass, confirm that nuclear effects not included in the oversimplified RFG model cannot be taken into account 
through a modification of $M_A$. In this context, it has to be pointed out that the need of a larger $M_A$ to reproduce the measured 
NCE $Q^2$-distribution is not likely to be ascribable to different nuclear effects in the CCQE and NCE channels. In fact, the ratio between the   
$Q^2$-distributions obtained from the RFG model and the spectral function approach, providing a measure of the effects of nuclear 
dynamics, turns out to be nearly identical for CCQE and NCE. The difference does not exceed 2\% over the whole $Q^2$ range.
 
Our analysis also shows that the strange quark contribution to the cross section of nuclei with equal number of protons and neutrons is vanishingly small. 
As a consequence, the possibility of  improving the agreement between MC simulations and Carbon data adjusting the value of  $\Delta s$ appears to be ruled out.
 
The authors of Ref.\cite{newpar} argued that the disagreement between theory and MiniBooNE CCQE data may be due to the uncertainties associated 
with the flux average procedure, as the resulting  cross section at fixed energy and scattering angle of the outgoing muon picks up contributions from 
different kinematical regions, where different reaction mechanisms are known to be dominant.

This uncertainty also affects the flux averaged NCE differential cross section, which is given in bins of  {\em reconstructed} $Q^2$ \cite{BooNENCE}, 
defined as
\beq
Q^2_{rec} = 2 M T = 2 M \sum_i T_i \ ,
\eeq
where M is the nucleon mass and $T$ is the sum of the
kinetic energies of the final state nucleons.

In order to provide results that can be compared to data in a meaningful fashion, theoretical models of neutrino-nucleus interactions must be based 
on a consistent description of the broad kinematical range corresponding to the relevant neutrino
energies. In the quasi elastic sector, this amounts to taking into account, besides single-nucleon knock out,  multi-nucleon knock out as well as processes 
involving the nuclear two-body currents, whose contribution is expected to be significant \cite{barbaro}.


\end{document}